\documentclass[preprint2]{aastex631} %%%%double-column, single-spaced document
\usepackage{graphicx}
\usepackage{amsmath}

\received{June 03, 2024} %\received{January 1, 2018}
\revised{September 16, 2024} %\revised{January 7, 2018}
\accepted{September 18, 2024} %\accepted{\today}
\newcommand{\p}[1]{{\color{magenta}{#1}}}

\newcommand{\PC}{{\sc Pencil Code}~}

\shorttitle{Simulating the formation and eruption of flux rope}
\shortauthors{Vemareddy}

\begin{document}
\title{Simulating the formation and eruption of flux rope by magneto-friction model driven by time-dependent electric fields}

%\correspondingauthor{P.~Vemareddy}
%\email{vemareddy@iiap.res.in}

\author{P.~Vemareddy}    % [0000-0003-4433-8823]
\affil{Indian Institute of Astrophysics, II Block, Koramangala, Bengaluru-560 034, India}
  
\begin{abstract}
Aiming to capture the formation and eruption of flux ropes (FRs) in the source active regions (ARs), we simulate the coronal magnetic field evolution of the AR 11429 employing the time-dependent magneto-friction model (TMF). The initial field is driven by electric fields that are derived from time-sequence photospheric vector magnetic field observations by invoking ad-hoc assumptions. The simulated magnetic structure evolves from potential to twisted fields over the course of two days, followed by rise motion in the later evolution, depicting the formation of FR and its slow eruption later. The magnetic configuration resembles an inverse S-sigmoidal structure, composed of a potential field enveloping the inverse J-shaped fields that are shared past one another and a low lying twisted field along the major PIL. To compare with observations, proxy emission maps based on averaged current density along the field lines are generated from the simulated field. These emission maps exhibit a remarkable one-to-one correspondence with the spatial characteristics in coronal EUV images, especially the filament-trace supported by the twisted magnetic field in the south-west subregion. Further, the topological analysis of the simulated field reveals the co-spatial flare ribbons with the quasi-separatrix layers, which is consistent with the standard flare models; therefore, the extent of the twist and orientation of the erupting FR is indicated to be the real scenario in this case. The TMF model simulates the coronal field evolution, correctly capturing the formation of the FR in the observed time scale and the twisted field generated from these simulations serve as the initial condition for the full MHD simulations.
\end{abstract}

\section{Introduction}
%\linenumbers
Coronal mass ejections and flares are the most spectacular and dangerous manifestations of solar activity. They are believed to be powered by the release of vast amounts of magnetic energy \citep{Wiegelmann2017_CorMagFld}. These events occur from the corona of the active regions (ARs) where intense magnetic fields are present. Coronal imaging observations in extreme ultraviolet (EUV) and soft X-ray wavebands reveal that the corona has a complex and evolving magnetic structure that is continuously driven by ubiquitous plasma motions at the photospheric surface. Because the solar corona satisfies the low-$\beta$ plasma condition, magnetic fields play a fundamental role in these energetic events \citep{shibata2011} and therefore understanding the structure and evolution of the coronal magnetic fields is crucial to comprehending the genesis of the space-whether affecting solar events.

High-resolution and high-cadence magnetic field measurements of the photospheric surface are regularly available from the ground (e.g., \citealt{Harvey1996_GONG,Keller2003_SOLICVSM}) and space (e.g., \citealt{Scherrer1995_MDI, Scherrer2012_HMIInstr}) instruments but the coronal magnetic field measurements are difficult to make due to tenuous plasma conditions. There are occasionally direct measurements of the coronal field available off the solar limb (e.g., \citealt{LinH2004_CorMagMeas, Tomczyk2008_CoMP, SiR2020_SpecMagMeas, LandiE2020_HinodeMagMeas}. However, these measurements frequently suffer from line-of-sight confusion and inadequate spatial or temporal resolution, making accurate interpretation difficult. While coronal magnetometry is still being developed \citep{GibsonSE2016_FORWARD, Raouafi2016_DiagCorMag}, a significant research has been focused on numerical modelling of the coronal magnetic field using photospheric vector magnetograms (e.g., \citealt{Sakurai1989_MagField_ext, gary1989, Schrijver2008_nlff}) in the past two decades.

Non-linear force-free field (NLFFF) models have been used to reconstruct the three-dimensional (3D) coronal magnetic field using the observed photospheric magnetograms (e.g.,\citealt{derosa2009, Regnier2013_MagFldExt, Wiegelmann2017_CorMagFld}). These models' main assumption is that the Lorentz force vanishes in the corona, allowing field-aligned currents to deform the field geometry, thereby offering estimates of magnetic non-potential quantities such as relative magnetic helicity and free magnetic energy. The NLFFF models have been employed to study the coronal magnetic structure constituting topological features such as null points, twisted flux tubes, flux ropes (e.g., \citealt{Regnier2004_3DMagConf, mason2009, vemareddy2014_Quasi_Stat, Vemareddy2018a}). The extrapolated magnetic structure is validated by comparing the chosen model field lines with the observed magnetic structure in e.g. EUV or soft X-ray images. These models are successful in yielding comparable structures resembling the observed ones in most of the cases \citep{Schrijver2008_nlff, derosa2009}). It is important to note that the extrapolated magnetic fields represent static fields in equilibrium where the dynamic evolution is missing. A useful technique for producing the dynamical evolution of the magnetic fields in the AR corona is to use magnetohydrodynamic (MHD) simulations (e.g., \citealt{Mikic1999_mhd_globalcor, Inoue2016_MHDmod, Toriumi2020, Jiang2022_DataDriven_CorMagFld}). These simulations involve advancing the full MHD equations in time and are physically realistic to capture information about plasma flow, density and temperature in addition to the magnetic field (e.g., \citealt{Mikic1999_mhd_globalcor, GN02}) during various physical processes such as flare reconnection in the computational domain. Even though these full MHD simulations are physically realistic, they are computationally expensive for domains of the AR size at meaningful resolutions, particularly when examining long-term evolution of the order of a few days.

Magneto-friction (MF; \citealt{Yang1986_MagFric}) method is another approach to construct NLFFF based on the induction equation alone (e.g. \citealt{valori2005,bobra2008} ). In this approach, the inductive velocity is proportional to the Lorentz force, which is relaxed to force-free state over time. The MF model can be driven by the time-series of lower boundary observations to simulate the dynamical evolution of the coronal magnetic field \citep{vanBallegooijen2000_FilChan, Mackay2011_ModDis}. In wake of availability of time series photospheric magnetic field observations at high cadence and high resolution from the Helioseismic and Magnetic Imager (HMI; \citealt{schou2012}) onboard Solar Dynamics Observatory (SDO), time-dependent data-driven MF simulations have become a useful technique to study the long term evolution of the ARs. 

Utilizing the line-of-sight component of magnetic fields,  inductive components of electric fields are derived by inversion methods developed by \citet{Fisher2010_EleEst}. The inductive electric fields have been used to drive the coronal field \citep{Mackay2011_ModDis, Gibb2014_mf}. In these simulations, the energy injection depends on the LOS component alone and the information of the horizontal magnetic field is missing to supply enough energy input into the coronal field. To account for additional energy injection, \citet{Cheung2012_MF} supplemented the non-inductive components to the inductive ones by employing ad hoc assumptions and found that the non-inductive electric fields are crucial for significant energy injection that is essential to form energized magnetic structures such as flux ropes in the corona. Further such simulations have been employed to study the formation of the helical jets, and twisted flux ropes in the ARs over days \citep{Cheung2015_helicaljets, Pomoell2019_EleFld}. In this article, using the time-dependent MF model, we simulate the coronal field evolution in AR 11429 by driving the initial field with electric fields that are derived from a time sequence of observed vector magnetic fields. The numerical setup and methodology are described in Section~\ref{SimSetup}. A detailed analysis of the simulated magnetic fields in comparison with coronal observations is presented in section~\ref{Res} and summary of the results is outlined in Section~\ref{Summ}.

\section{Simulation model and setup}
\label{SimSetup}
\subsection{Coronal field model}
We simulate the coronal magnetic field evolution by magneto-friction (MF) method \citep{Yang1986_MagFric}, where the magnetic field evolves in response to the photospheric flux motions through the non-ideal induction equation given by   
\begin{equation}
\frac{\partial \mathbf{A}}{\partial t}= \mathbf{v}_{\rm MF}\times\mathbf{B}-\eta \mu_0 \mathbf{J}
\label{eq_Ind}
\end{equation}
where $\mathbf{v}_{\rm MF}$ is the magnetic frictional plasma velocity, $\mathbf{A}$ is the vector potential relating the magnetic field as $\mathbf{B}=\nabla\times\mathbf{A}$. The second term is to accommodate dissipation in the corona due to electric currents $\mathbf{J}=\nabla\times\mathbf{B}/\mu_0$. The magnetic diffusivity can be chosen a typical value of $\eta=1\times10^{8}\,m^2/s$ to be able to run the simulation stable. Here the induction equation is solved in terms of $\mathbf{A}$ to ensure that $\nabla\cdot\mathbf{B}=0$ without additional divergence-free schemes. From the assumption of static magnetic fields, the MF velocity is given by
\begin{equation}
\mathbf{v}_{\rm MF}=\frac{1}{\nu}\frac{\mu_0\, \mathbf{J}\times\mathbf{B}}{B^2}
\label{eq_v_mf}
\end{equation}
where $\nu$ is the magneto-frictional coefficient that controls the speed of the relaxation process and $\mu_0$ is the magnetic permeability in vacuum.  As suggested in \citet{Cheung2012_MF} a height-dependent form of the frictional coefficient $\nu$ is given by
\begin{equation}
\frac{1}{\nu} = \frac{1}{\nu_0}(1-e^{-z/L}),
\label{eq_nuprof}
\end{equation}
where $\nu_0$ is set around $35\times10^{-12}$ s\,m$^{-2}$ and $z$ is the height above the bottom boundary, and $L$ is the chosen as 15 Mm. This form of frictional coefficient gives MF velocities a smooth transition to zero towards bottom boundary $z=0$.

Along with a special driver module, \citet{Vemareddy2024_DataDriven} first implemented the MF model in \PC \citep{PC_collab2021} and tested the evolution of coronal magnetic field driven by observed photospheric magnetic fields. The \PC is a highly modular physics-oriented simulation code that can be adapted to a wide range of applications. It is a finite-difference code using sixth order in space and third-order in time differentiation schemes.

\subsection{Initial Conditions}
We use potential field (PF) model \citep{gary1989}) as the initial field for the simulation. The PF is constructed from the normal component ($B_n \to B_z$) of the observed vector magnetic field of the AR 11429 at 06:00 UT on March 2012. As the \PC operates on the vector potential ($\mathbf{A}$) rather than magnetic field, the field divergence is being satisfied without the need for extra schemes. The vector potential $\mathbf{A}_p$ of the PF is computed by imposing coulomb gauge condition ($\nabla \cdot \mathbf{A}_p=0$) and has vanishing normal component ($\mathbf{A}_p \cdot \hat{n}=0$) at the bottom boundary (e.g., \citealt{DeVore2000}). With these constraints, the components of $\mathbf{A}_p$ are computed as,

\begin{eqnarray}
A_{{\rm p},x}(x,y,z)&=&{\rm FT}^{-1}\left[ \frac{ik_y}{k^2}\,{\rm FT} [B_z(x,y)]\,e^{-kz} \right] \\
A_{{\rm p},y}(x,y,z)&=&{\rm FT}^{-1}\left[ \frac{-ik_x}{k^2}\,{\rm FT} [B_z(x,y)]\,e^{-k z}\right] 
\end{eqnarray}
where $k_x$ and $k_y$ are wave vectors along the x and y directions, respectively, and $k=\sqrt{k_x^2+k_y^2}$. Here, FT refers to the 2D Fourier transform operation on the observed $B_z(x,y)$ and $FT^{-1}$ refers to inverse Fourier transform. The observed $B_z$ is adjusted to satisfy flux balance condition and is inserted in a larger area by padding the dimensions. The $\mathbf{A}_p$ is constructed on a uniform Cartesian computational grid of $192\times192\times120$ representing the AR coronal volume of physical dimensions $ 280\times280\times175 $ Mm$^3$. 

\subsection{Time-dependent boundary conditions}
In the data-driven MF models, the initial magnetic field at the start time of the simulation is driven by magnetic field observations of the photospheric boundary that represents bottom of the computational domain \citep{Mackay2011_ModDis, Yardley2018_MF_11437}. We obtained the ambiguity resolved HMI vector magnetic field components which are served as \texttt{hmi.sharp.cea\_720s} data product in spherical coordinates. These are approximated to Cartesian components as $B_x \to B_\phi$, B$_y\to-B_\theta$ and $B_z\to B_r$ \citep{bobra2014} and are smoothed with a Gaussian width of 3 pixels both spatially and temporally. Also, the flux balance condition is applied to the $B_z$ component. In order to reduce computation burden, the field components are rebinned by a factor four such that each pixel corresponds 2" on the computation grid. Driving the initial 3D field with photospheric magnetic fields leads to mild to moderate non-potential fields, which may not represent the twisted core field of the ARs \citep{Yardley2018_MF_11437}. For complex active regions such as 11429, the observations have insufficient influx of magnetic energy and helicity, as a result the formation of low lying twisted flux in the simulations has not been possible \citep{Vemareddy2024_DataDriven}. In this work, the initial field is driven by the electric fields ($\mathbf{E}(x,y,z=0,t)$), which are derived from time sequence vector magnetic field observations of the AR \citep{Fisher2010_EleEst}. The electric field has two components: viz inductive field ($\mathbf{E}_I$) and non-inductive component defined by gradient of a scalar $\psi$
\begin{equation}
\mathbf{E} = \mathbf{E}_I- \nabla \psi
\end{equation}
The inductive component is obtained by solving the Faradays' law of induction equation, $\partial_t \mathbf{B}= -\nabla \times \mathbf{E}_I$, which involves decomposing the poloidal-toroidal decomposition of the time derivative of the vector magnetic fields \citep{Fisher2010_EleEst}. The non-inductive component requires a functional form for $\psi$ which is unknown and therefore invokes the following three different assumptions \citep{Cheung2012_MF, Cheung2015_helicaljets}:
\begin{eqnarray}
\nabla^2\psi&=&0 \label{ass_0}\\
\nabla^2 \psi &=& U \left ( \nabla \times \textbf{B} \right )_z \label{ass_u}\\
\nabla^2 \psi &=& \Omega B_z \label{ass_o}
\end{eqnarray}
where $\Omega$ and $U$ are the free scalar parameters referring to the extent to which the axisymmetric vertical flux tube rotates and emerges, respectively. The first assumption is equivalent to zero contribution of non-inductive components, whereas the latter two cases are suggested by \citet{Cheung2012_MF, Cheung2015_helicaljets} in order to ensure sufficient injection of magnetic helicity and free magnetic energy. The above chosen functional forms are linked to photospheric magnetic field observations and their horizontal spatial derivatives, therefore the non-inductive part of $\mathbf{E}$ has horizontal components (x and y) that are being added to inductive part of 
$\mathbf{E}$. This form for non-inductive $\mathbf{E}$ have vanishing horizontal divergence, as a result satisfies the induction equation such that $\partial B_z/\partial t=-\hat{z}\cdot(\nabla\times\mathbf{E}_h)$  without modifying the $B_z$.

\begin{figure*}[!ht]
\centering
\includegraphics[width=.99\textwidth,clip=]{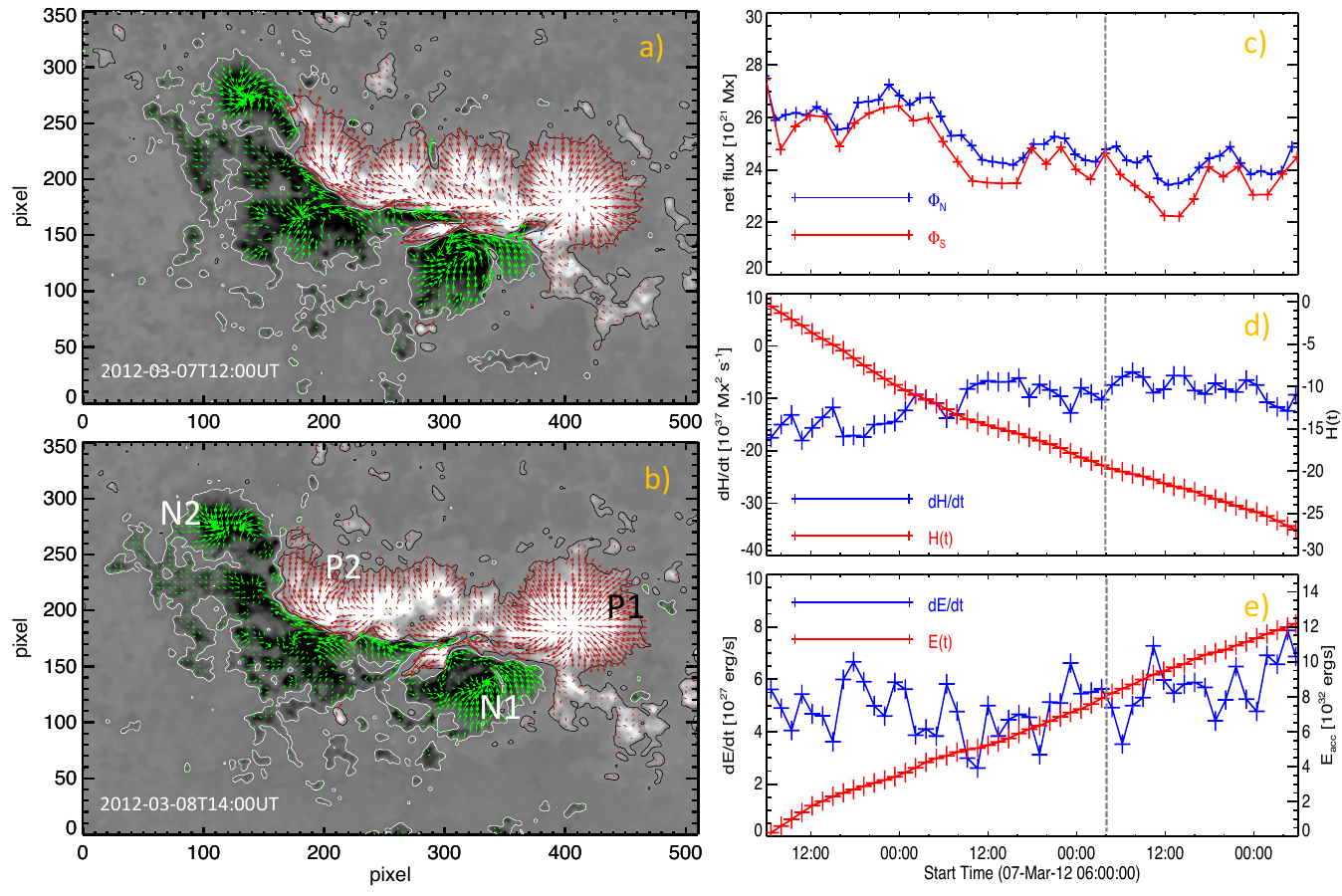}
\caption{(a)–(b)) Vector magnetic field measurements of the AR 11429 at two different times in the evolution. Horizontal field vectors (arrows) are overplotted on $B_z$ map with labelled polarity regions. The axis units are in pixels of 0".5.  (c-e) time evolution of net magnetic flux, helicity injection rate ($dH/dt$), and energy injection rate ($dE/dt$). Vertical dotted line refers to the eruption time at 03:44 UT on March 9. }
\label{fig1}
\end{figure*}

The derived photospheric electric fields at an interval of 12 minutes, are being used as driver field $\mathbf{E}$ at $z=0$, instead of magnetic fields or vector potentials. At each instant of time, we add $\mathbf{E}$ at the bottom layer of the vector potential $\mathbf{A}$ in the computational domain and the time evolution of the magnetic field is determined by the time-integration of the Faraday's law
\begin{equation}
\frac{\partial \mathbf{A}}{\partial t} = -\mathbf{E}
\end{equation}
In the \PC, we developed a driver module to read the time-series of $\mathbf{E}$ at 12 minute intervals and interpolate them at the simulation time-step.

The lateral sides of simulation domain are specified with periodic boundary conditions and the top one is set to open boundary condition.

\section{Results}
\label{Res}
The AR 11429 was a successively erupting region during its disk passage \citep{DhakalS2020, Vemareddy2021_MagStru}. It produced three fast CMEs at an interval of about 2 days, which is a typical timescale for the buildup of magnetic energy by photospheric magnetic flux motions. As an example, the HMI observations of the photospheric vector magnetograms at two different times are displayed in Figure~\ref{fig1}(a-b). The AR has a long polarity inversion line (PIL) in the north-east (NE) subregion between P1 and N1 regions, in the south-west (SW) region between P2 and N2. These polarity regions exhibit shear motions, which can deform the magnetic field to align parallel to the local PIL. As a result, the horizontal vectors along the PIL are noticed to align parallel to the PIL as observed in the vector magnetograms. Both shear and converging motions are potential mechanisms to form a twisted magnetic flux at the core of the AR and then its further eruption \citep{amari2003a}.  Similar to the successive erupting AR 12371 \citep{Vemareddy2017_SuccHomol}, \citet{DhakalS2020} suggested that the shearing motion and magnetic flux cancellation of opposite fluxes were the dominant factors to the recurrent homologous eruptions from this AR and are associated with the filament lying along the SW PIL. Under these observed conditions of the magnetic fluxes, after the first eruption on March 7 at 00:24 UT, this study is aimed at simulating the coronal field evolution until the next eruption on March 9 at 03:53 UT and then capturing the formation of the twisted flux rope that could erupt. To have relaxed and approximate potential field configuration after the first eruption, the simulation starting time is chosen to be 6:00 UT on March 7.

The time evolution of the derived magnetic parameters of vector magnetic field observations of the AR 11429 are plotted in Figure~\ref{fig1}(c-e). The net flux ($ \Phi_{N/S}=\sum_{Pix}Bz_{N/S} dA $, where $dA$ is area of the pixel in north (N) or south (S) polarity ) exhibits decreasing evolution in time from $26\times10^{21}$ Mx to $24\times10^{21}$ Mx in both polarities, which is presumed to be due to the converging motion of fluxes leading to their cancellations. From the time sequence vector magnetic fields ($\mathbf{B}$), the velocity field ($\mathbf{V}$) is derived by \textit{differential affine velocity estimator} (DAVE4VM; \citealt{schuck2008}) and then compute the helicity injection rate through the photosphere as
\begin{equation}
{{\left. \frac{dH}{dt} \right|}_{S}}=2\int\limits_{S}{\left( {{\mathbf{A}}_{P}}\cdot {{\mathbf{B}}_{t}} \right){{\text{V}}_n}dS}-2\int\limits_{S}{\left( {{\mathbf{A}}_{P}}\cdot {{\mathbf{V}}_{t}} \right){{\text{B}}_{n}}dS}
\label{eq_dhdt}
\end{equation}
where $\mathbf{A}_p$ is the vector potential of the potential field $\mathbf{B}_p$, and the variables with subscripts $n$, $t$ refer to normal and tangential components. The flux motions are injecting a net negative helicity at an average rate of $9\times10^{37}$ Mx$^2\,s^{-1}$. At this rate, the corona has an accumulated net helicity of $20\times10^{42}$ Mx$^2$ over two days of evolution, which could be sufficient to launch a CME.

\begin{figure*}[!ht]
\centering
\includegraphics[width=.8\textwidth,clip=]{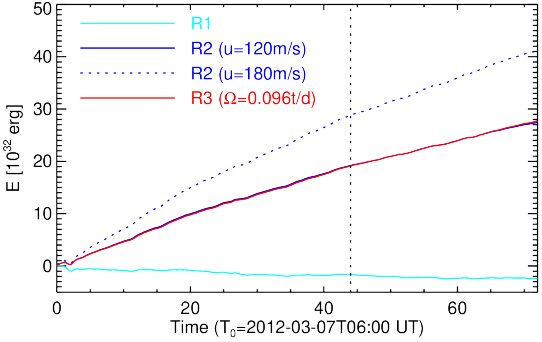}
\caption{Time-integrated Poynting flux from the AR derived from photospheric electric field. The run R1 is performed with an inductive electric field ($\mathbf{E}_I$), whereas R2, R3 are performed by electric fields, accounting non-inductive contributions based on ad-hoc assumptions, viz. equations~\ref{ass_u} and~\ref{ass_o} respectively. R2 is performed with $U=120\,m\,s^{-1}$ and $U=180\,m\,s^{-1}$. Vertical dotted line refers to the eruption time at 03:44 UT on March 9. }
\label{fig_exb}
\end{figure*}

Similarly, the energy injection rate (Poynting flux) is calculated as
\begin{equation}
{{\left. \frac{dE}{dt} \right|}_{S}}=\frac{1}{4\pi }\int\limits_{S}{B_{t}^{2}{{V}_{n}}dS-\frac{1}{4\pi }}\int\limits_{S}{\left( {{\mathbf{B}}_{t}}\cdot {{\mathbf{V}}_{t}} \right){{B}_{n}}dS}
\label{eq_dedt}
\end{equation}
which is time integrated to evaluate the accumulated energy over a time period T as 
\begin{equation}
E_{acc}=\int_{0}^{T}\frac{dE}{dt} dt
\label{eq_e_acc}
\end{equation}
As can be noted from Figure~\ref{fig1}e, the observed energy injection leads to an accumulated energy of $8.4\times10^{32}$ ergs, a energy budget for M-class flares.

\begin{figure*}[!ht]
    \centering
    \includegraphics[width=.7\textwidth,clip=]{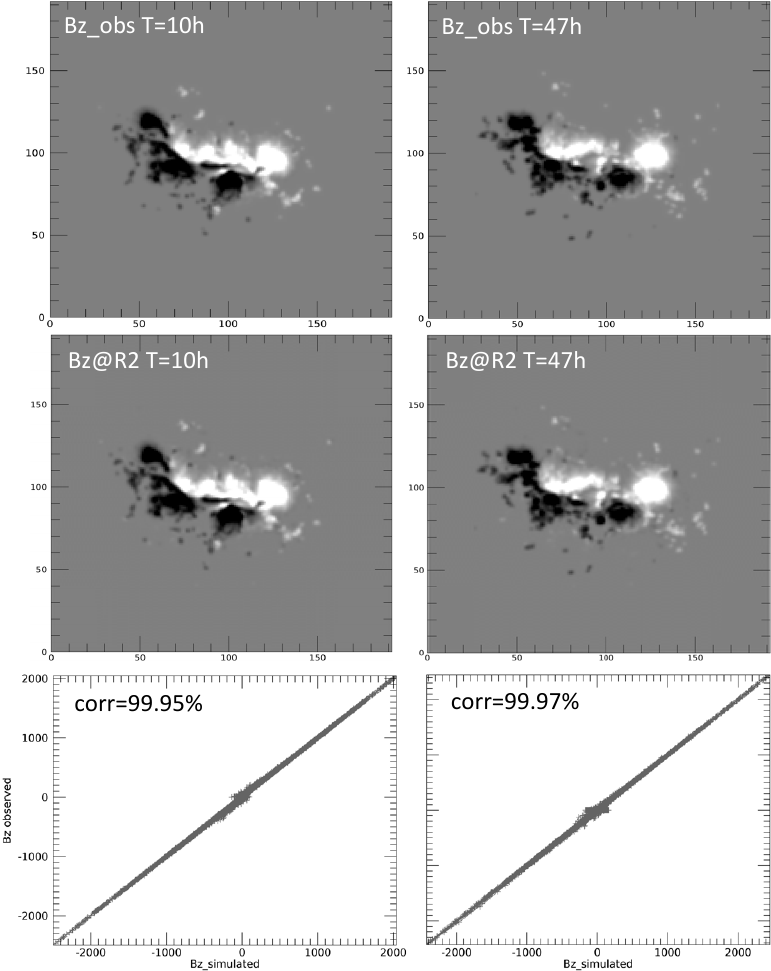}
    \caption{Reproducibility of observed Bz at the photosphere with MF simulation driven by $\mathbf{E}$-field. {\bf Top row:} Observed $B_z$ at 10h and 47h from start of the simulation (i.e., 2012-03-07T06:00 UT) {\bf middle row:} $B_z|_{z=0}$ of R2 at 10h and 47h  which have one-to-one spatial correlation with the observations. Note that these maps are scaled with $\pm900$ G, {\bf bottom row:} scatter plots of observed and simulated Bz showing correlation of 99.95\%. }
    \label{fig_bz_corr}
\end{figure*}

\subsection{Input Poynting flux for simulations}
We drive the initial PF with the time-dependent electric field derived from the ad hoc assumptions given in equations \ref{ass_0}-\ref{ass_o}. With the electric fields from the three assumptions, we perform essentially three simulation runs. The run R1 is based on the inductive electric field ($E_I$), whereas runs R2 and R3 are included with non-inductive contributions in order ensure sufficient injection of energy and helicity fluxes. For the later two runs, one has to specify appropriate values for free parameters $U$ and $\Omega$. As suggested by \citet{Pomoell2019_EleFld}, these values are constrained by comparing the Poynting flux deduced from the DAVE4VM vector velocity and magnetic field (Figure~\ref{fig1}e, and equation~\ref{ass_o}) with that of the derived electric field ($dE/dt=\int \mathbf{E} \times \mathbf{B} \cdot d\mathbf{A} $) from these assumptions \citep{Pomoell2019_EleFld}. 

We use the  values of $U=120$\,m\,s$^{-1}$ and $\Omega=0.096$ turns/day for R2 and R3, respectively. As shown in Figure~\ref{fig_exb}, these values give a time-integrated poynting flux of $18\times10^{32}$ ergs till the eruption time (vertical dotted line), which is roughly a factor of two higher when compared with the observed one (Figure~\ref{fig1}e). This difference is admitted for the reasons that i) the AR could be in a non-potential state even after the first eruption, ii) the Poynting flux derived from DAVE4VM may be underestimated because the velocities represent averaged values over an apodising window typically $19\times19$ pixel$^2$, iii) observational sensitivity of magnetic field measurements. With these existing difficulties, the chosen values of U and $\Omega$ are broadly constrained with an uncertainty of upto 40\%. It should be noted that the coronal free energy is small (about 18 times) and the formation of the twisted flux is quite implausible with $\mathbf{E}_I$ alone, therefore higher values of these free parameters are deemed in order to pump sufficient free magnetic energy to generate magnetic structure that is comparable to observations

\begin{figure*}[!ht]
\centering
\includegraphics[width=.95\textwidth,clip=]{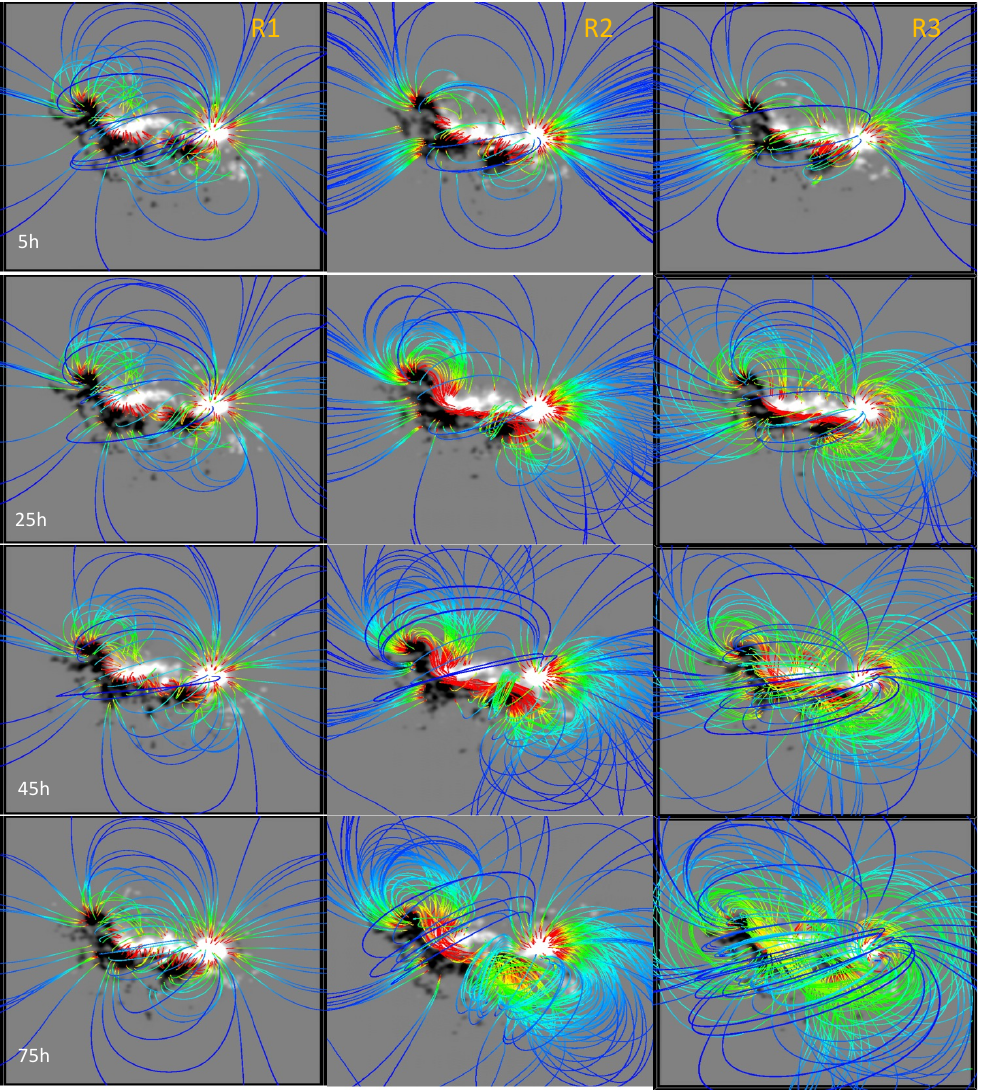}
\caption{Snapshots of magnetic structure at different epochs of the simulation: R1 (first column), R2 (second column) and R3 (third column). Magnetic field lines are colored by their field strengths and the background image is the normal magnetic field $B_z$ at the bottom of the computational domain ($z = 0$). In R2 and R3, the magnetic field evolution captures the formation of the twisted flux rope along the PIL. The image sequence of all three simulation runs at 1 hour interval is attached as an animation. }
\label{fig_magst}
\end{figure*}

\subsection{Evolution of the simulated magnetic structure}
The above said three types of MF simulations are performed by driving the initial PF with a time dependent bottom boundary electric field for 72 hours duration starting from 6:00 UT on March 7, 2012, and further 24 hours without changing the lower boundary. 

As explained earlier, the $\mathbf{E}$-field satisfies $\partial B_z/\partial t=-\hat{z}\cdot(\nabla\times\mathbf{E}_h)$ reproducing the observed evolution of $B_z$ self-consistently during the simulations. To check this , in Figure~\ref{fig_bz_corr}, we compare observed and simulated $B_z$ at $z=0$ plane at 10h and 47h time instants of simulation R2. As can be noticed that simulated $B_z$ has a high degree of spatial correlation including small magnetic elements in the vicinity of sunspot regions. The scatter plots refer to a correlation of $>99.95\%$ which implies the observed evolution of photospheric magnetic fields during the simulation. However, such a correlation is not true for horizontal components of $\mathbf{B}$ which are scaled by the free-parameters in non-inductive contribution of $\mathbf{E}$ in order to inject additional free-energy and helicity.

\begin{figure*}[!ht]
    \centering
    \includegraphics[width=.65\textwidth,clip=]{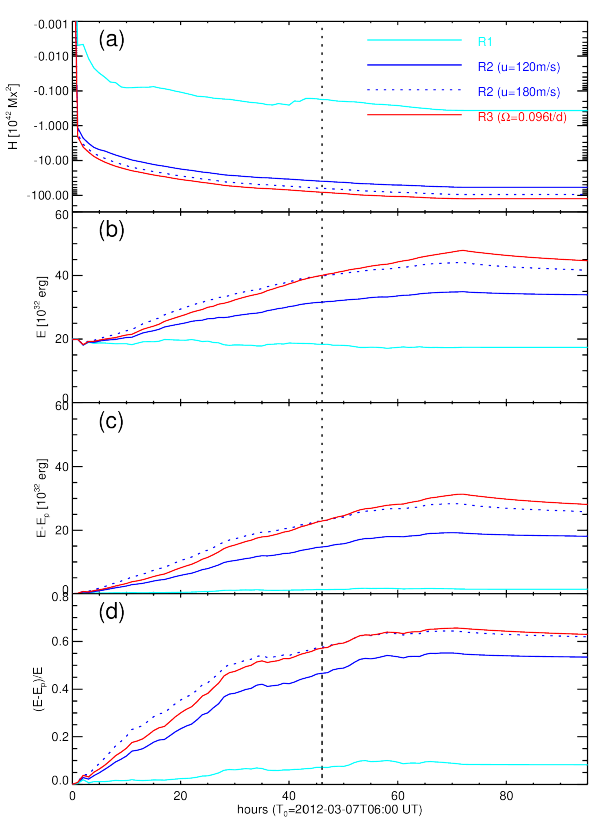}
    \caption{Time evolution of energy and helicity parameters in the computational domain. a) total magnetic helicity, b) total magnetic energy, c) free magnetic energy, and d) fractional free energy with respect to the total energy. Cyan, blue, and red color curves refer to runs R1, R2, and R3 respectively. Vertical dotted line refers to the eruption time at 03:44 UT on March 9, at which time the free-energy fraction is 7\%, 47\%, 57\% for R1, R2 and R3 respectively.} 
    \label{fig_mets}
\end{figure*}

\begin{figure*}[!ht]
\centering
\includegraphics[width=.95\textwidth]{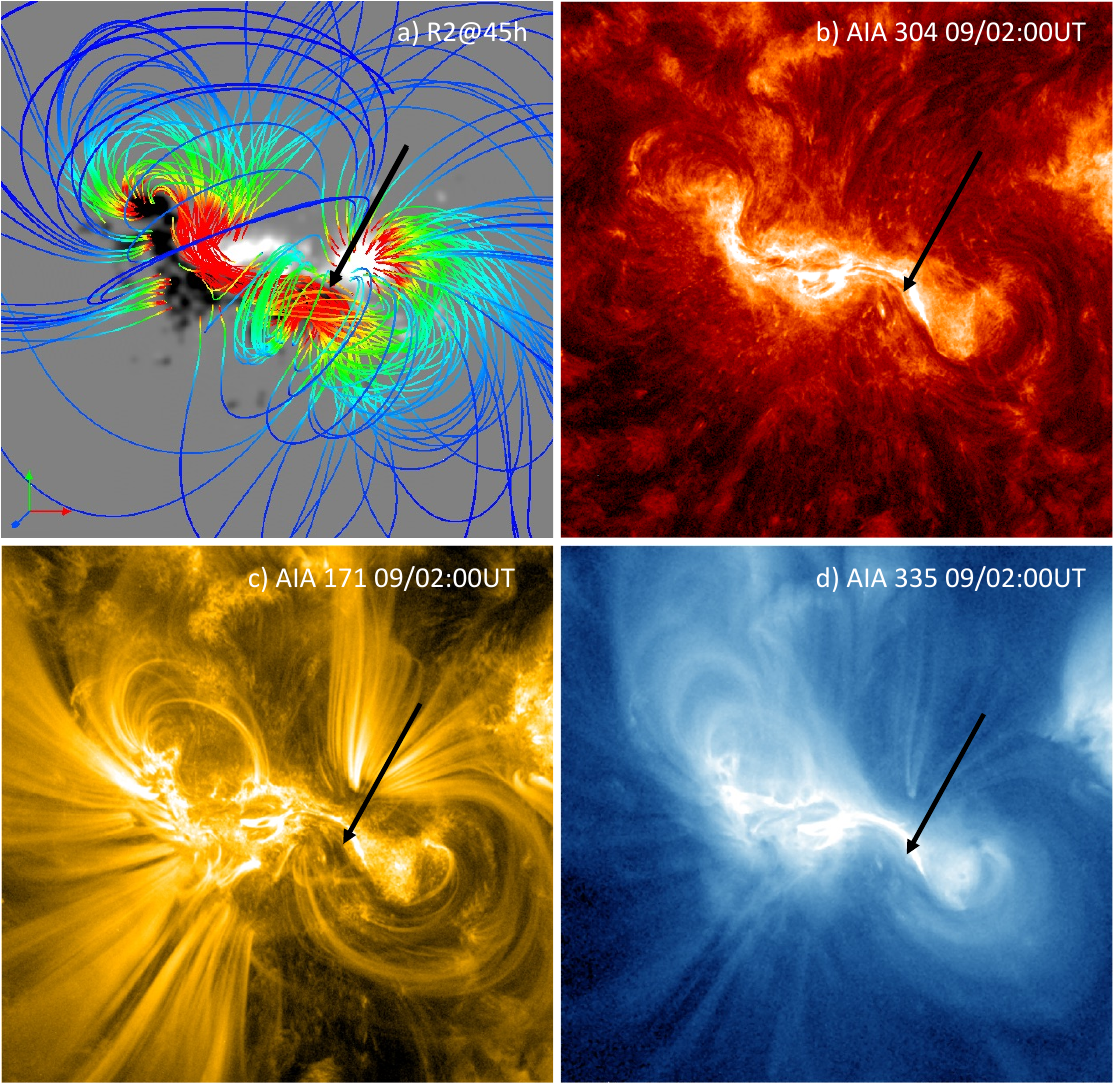}
\caption{Comparison of pre-eruptive magnetic structure of R2 ($U=180\,m/s$) with the coronal plasma tracers in AIA images of the AR 11429. {\bf a)} top view of rendered magnetic structure at 45th hour of simulation R2. Background image is $B_z$ distribution at $z=0$. {\bf b-c)} AIA 304, 171, and 335~\AA~images at 2:00UT on March 9. The morphology in these images depicts the structured corona of twisted flux rope that resembles the simulated magnetic field. Black arrow points to the eruptive filament manifested by twisted field in SW region.}
\label{fig_magst_aia}
\end{figure*}

Figure~\ref{fig_magst} presents the snapshots of the simulated magnetic structure of all three runs. Field lines are traced from the foot point locations where the horizontal magnetic field and total electric current are strong. From R1, one can see that the initial potential field does not change significantly over several hours of evolution as the boundary electric field has inductive components derived from $B_z$ components alone. In the runs R2 and R3, the initial magnetic structure evolved progressively to a highly twisted structure at the core of the AR surrounded by a less sheared field which is evidently noticed in the snapshots at 25th hour. Further evolution till 45th hour, leads to transforming neighbouring less sheared fields into twisted fields akin to a coherent flux rope along the PIL. Following this, the flux rope and the surrounding field grow and appear to rise in height, which is a typical signature of initiation of the eruption. In the presence of magnetic diffusion and supply of magnetic helicity through a time-dependent bottom boundary, the MF relaxation method thus captures the formation and eruption of MFR in the AR. Using the other codes, earlier studies (eg., \citealt{Cheung2012_MF, Pomoell2019_EleFld}) reported the simulated coronal field evolution capturing the flux rope formation in the emerging ARs. 

It is worth pointing out that the electric fields in R2 are derived with $J_z$ distribution at the photosphere, as a result, one can expect a spatially varying twist about the PIL scaled by the free parameter $U$. And in R3, the non-inductive contributions to the derived electric fields are based on $B_z$ distribution scaled by the $\Omega$, thus the spatial variation of magnetic twist along the PIL and its vicinity do not reflect the same observations as those of electric fields in R2. Given this fact, the simulated 3D magnetic structure in R2 better reflects the coronal EUV observations than that in R3, as will be convinced from the following comparative analysis with the observations. 

\begin{figure*}[!ht]
\centering
\includegraphics[width=.95\textwidth]{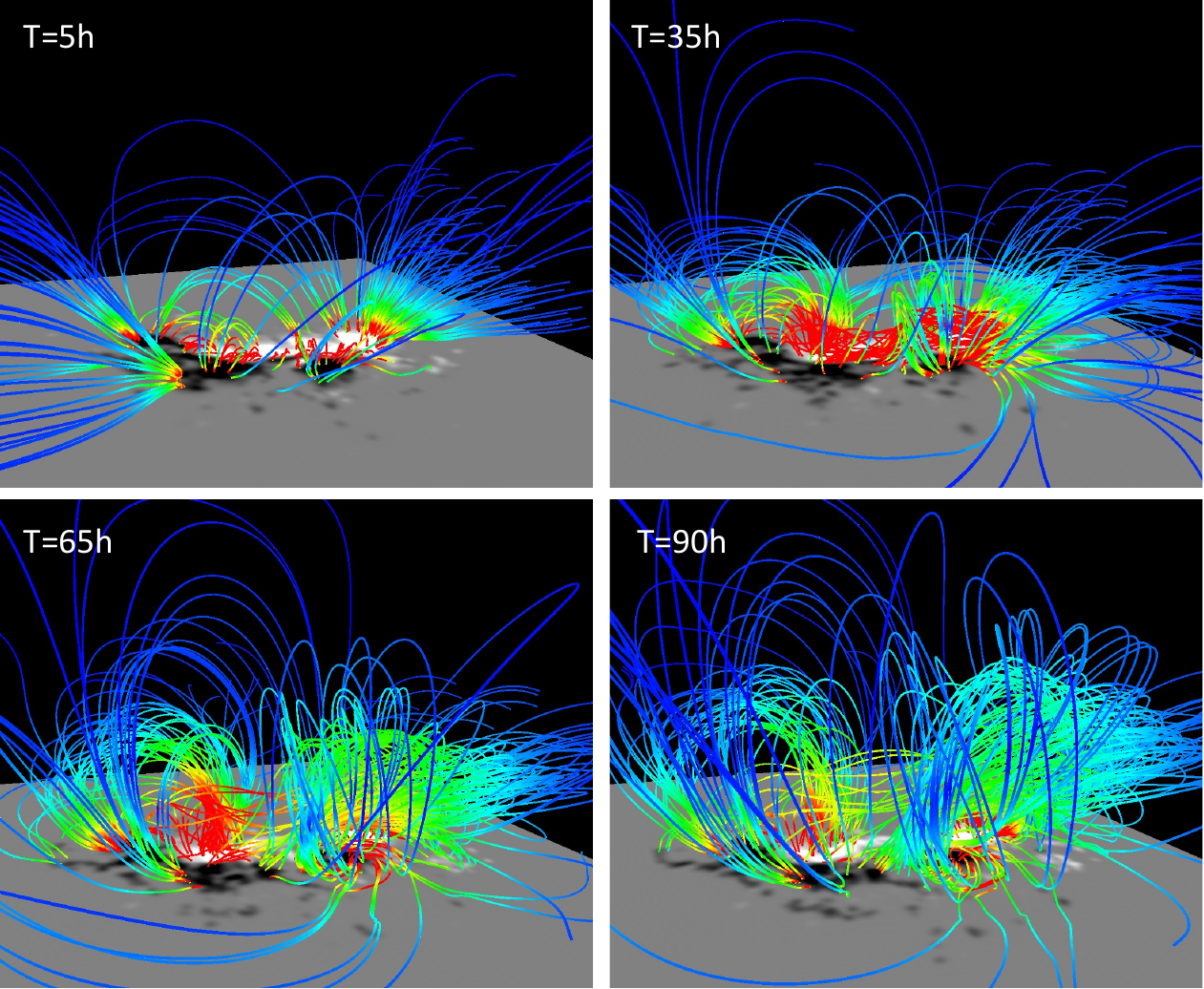}
\caption{Perspective view of the rendered magnetic structure of R2 at 5, 35, 65 and 90 hrs, respectively. From the initial potential field, the structure evolves to sheared field consisting of inverse J-shaped field lines surrounding the low lying twisted core field along the PIL. In a time scale of two days, a well developed twisted flux rope forms, which then runs into slow explosive stage in the corona. In order to comprehend the simulated evolution of this run, the image sequence of 96 hour duration at 1 hour interval is attached as an animation. Note that the filament embedded twisted structure in the SW region rises in height progressively.}

%From the initial potential field, magnetic flux rope forms in a time scale of two days consistent with observations. {\bf column 3-4}: AIA 304 and 171~\AA ~images at different times. The morphology in these images depicts the structured corona of twisted flux rope that resembles the simulated magnetic field.   
\label{fig_R2_persp}
\end{figure*}

The time evolution of computed relative magnetic helicity (H) and magnetic energy in the simulated domain are plotted in Figure~\ref{fig_mets}. In the MF relaxation of R1, the volume helicity (H) increases in magnitude from $-4\times10^{-9}$ Mx$^2$ to  $-0.37\times10^{42}$ Mx$^2$. This is solely due to shear motions of fluxes along PIL. Although there is no appreciable increase of total magnetic energy (E), the free magnetic energy ($E-E_p$) increases continuously, and it is $1.31\times10^{32}$ ergs at the time of the eruption. During the simulation time of R1, the fractional free energy increases upto 10\%. 

On the other hand, the H accumulates in the corona continuously (blue, red curves) and amounts to $-39.2\times10^{42}$ Mx$^2$, $-81.6\times10^{42}$ Mx$^2$ for R2 and R3 respectively. Similarly, the total and free magnetic energies exhibit increasing behaviour as the coronal magnetic field becomes twisted.  In these runs, the fractional free energy corresponds to 47\% (R2), 57\% (R3) at the time of the eruption, which is probably significant enough to initiate the rise motion of the flux rope. Numerical simulations by \citet{amari2003a} reported a fractional free energy of 5\% to initiate the eruption, while TMF simulations  by \citet{Pomoell2019_EleFld} refers to 14-50\% of fractional free energy that resulted in erupting flux rope, therefore, our results are more aligned with later one as the simulation setup is same. It is important to note that the coronal estimates of helicity ($H$) for R3 is higher by a factor of two than that for R2, even though the same amounts of energy fluxes are injected for these runs. As discussed earlier, the higher value of $H$ is due to uniform distribution of twist nature arising from equation~\ref{ass_u}, correspondingly, the coronal field is highly twisted in R3.

\begin{figure*}[!ht]
    \centering
    \includegraphics[width=0.75\textwidth]{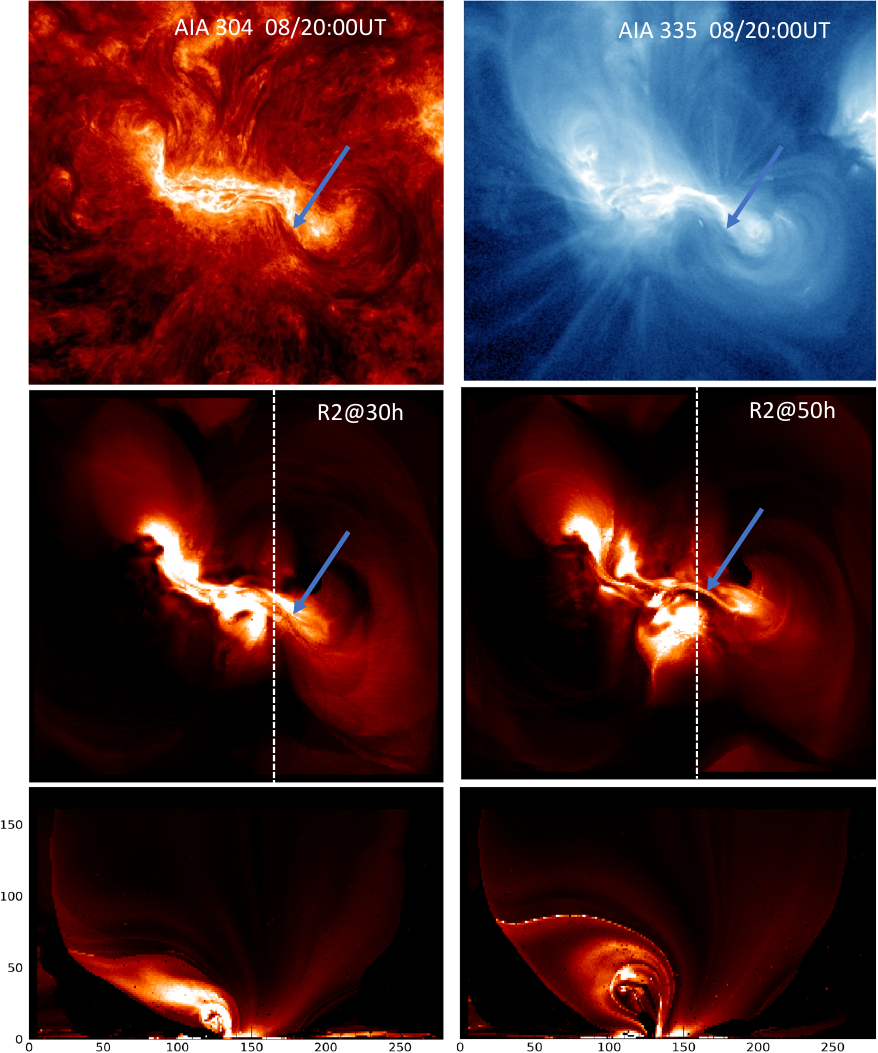}
    \caption{Comparison of proxy emission maps with the coronal EUV images. {\bf top row:} AIA 304~\AA~observations of the AR 11429 on two different days. {\bf middle row:} Proxy emission maps  synthesized from simulated magnetic field of R2 at the time instances of 30hr and 50hr. A visual inspection of these maps suggests that the spatial features presented in the AIA 304 and 335~\AA~images have a striking morphological similarity, especially low lying filament feature (pointed by arrow) and high lying coronal sigmoidal loops. White dashed line refers to position of vertical slice plane {\bf bottom row:} proxy emission in the vertical slice plane captures the twisted flux rope and its upward motion in time. Axes units are in Mm.}
    \label{fig_bemiss}
\end{figure*}

\subsection{Comparing the simulated magnetic structure with Coronal Observations}
We performed another simulation of R2 driven by electric fields with free parameter $U=180\,m/s$. This run is to understand the flux rope formation time scale and rise motion during a given time evolution. From the comparison, the twisted flux formation in this run happened early and later its rise motion to higher heights. Figure~\ref{fig_magst_aia}a displays the rendered magnetic structure of R2 ($U=180\,m/s$) with $B_z$ as background image. The modelled magnetic structure consisted of a low-lying, twisted core field along the PIL overlaid by potential arcades and J-shaped field lines (lobes or elbow field lines) sheared past each other in SW and NE regions, which together manifest an inverse-S sigmoidal structure. Although the global AR magnetic configuration is inverse S-shaped, the magnetic field in the SW region alone mimics another inverse-S sigmoid. As shown in Figure~\ref{fig_magst_aia}(b-c), the magnetic structure qualitatively resembles the morphology of coronal features captured in AIA images. Owing to high temperatures, sigmoids are often seen in hot EUV channels. AIA 335~\AA~image ($T\approx2.5$MK) displays coronal sigmoid resembling the simulated magnetic structure and plasma loops as the tracers of magnetic field in AIA 171~\AA snapshot. The AIA 304~\AA~waveband is a cool channel ($T\approx5\times10^4$K) image exhibiting a filament channel in the SW region, which is plasma embedded in the dips of twisted field of the sigmoid configuration.

The simulated magnetic structure is rendered in perspective, as depicted in Figure~\ref{fig_R2_persp} at different times. The initial potential field becomes sheared progressively forming inverse J-shaped field lines surrounding the low lying twisted core field along the PIL. In a time scale of two days a well developed twisted flux builds up along the PIL. Thereafter the twisted magnetic structure ascends in height with time, especially the field in the SW region, from 40 Mm to 120 Mm (See the attached animation). This rise motion corroborates the observations of the eruption from the SW region of the AR \citep{DhakalS2020}. We emphasise that the rise motion does not turns into an eventual eruption like the observed onset of eruptions, rather it is a progressive in time even after driving stopped from 72 hours. Since the velocity is controlled by $\nu$ parameter, the TMF simulations implicitly lacks the short term dynamic evolution, although they capture the build up of free magnetic energy during slow evolution over long time.

Since the MF model does not include the thermodynamic evolution of the plasma, we cannot synthesize images of coronal emission in EUV or X-ray wavelengths. To produce synthetic maps of coronal loops similar to EUV images, \citet{Cheung2012_MF} introduced a method to generate proxy emissivities based on the modelled magnetic field alone. Proxy emissivities are derived with the values of square of the total current density ($J^2$) averaged along a magnetic field line. This technique is useful for a qualitative visualization of coronal loops in a magnetic model such as MF, however it is not a replacement for more sophisticated techniques that use the thermodynamic variables from MHD models (e.g., \citealt{Lionello2009,Hansteen2010}). Following their method, we computed the emissivity ($\varepsilon(x,y,z)$) at each grid point in the computational domain. This volume distribution of emissivity is integrated in the vertical direction to generate a 2D distribution, as seen in coronal EUV observations. Such proxy emission maps at two epochs are compared with the respective AIA 304, 335~\AA~images in Figure~\ref{fig_bemiss}. We can notice a striking morphological similarity of the modelled emission with the diffuse plasma emission of the sigmoid captured in 335~\AA image and trace of dark filament present in 304~\AA~images. This filament is regarded as flux rope in the models and it is expanding and rising in time as observed in the vertical cross section planes. 

\begin{figure*}[!ht]
\centering
\includegraphics[width=0.9\textwidth]{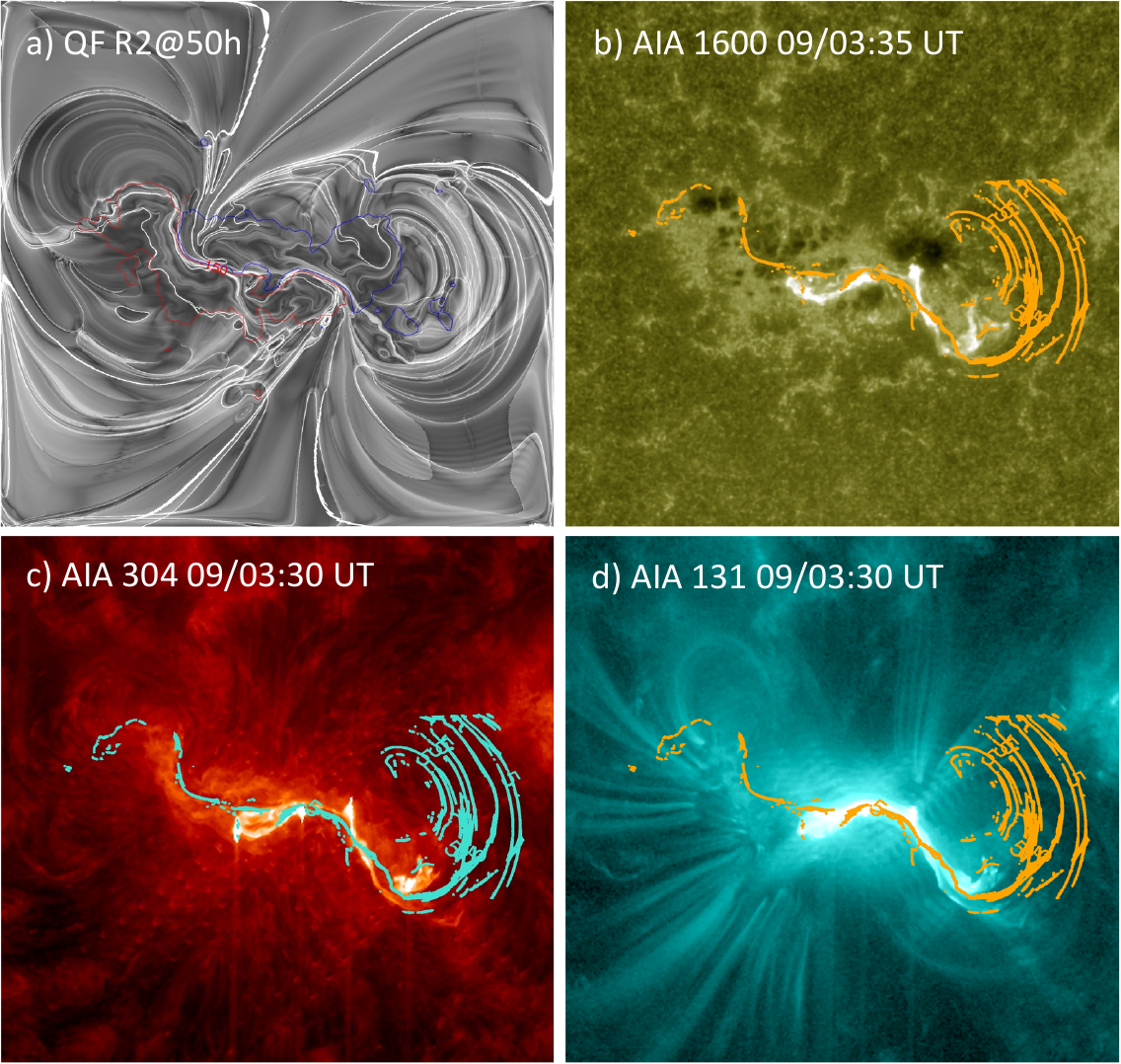} 
\caption{Comparison of QSLs with flare ribbons. {\bf a)} $Log(Q)$ map at $z=1.45$ Mm computed from the magnetic structure of R2 at 50 hr time instant. Contours of $B_z$ at $\pm150$ G are overdrawn (red/blue curves) and QSLs with large Q values are identified by intense white traces in strong field region. {\bf b-d)} AIA 1600, 304 and 131~\AA images overlaid with contours of ln(Q)=[5,6] (in cyan/orange color). Non-relevant QSLs are removed by applying mask on Q-map. Note the intense flare ribbon emission underneath the erupting flux rope is co-spatial with the QSL section in the SW subregion. 
}
\label{fig_qf_2d}
\end{figure*}

\subsection{Topological analysis of simulated magnetic structure}
For a topological study, we analyse the quasi separatrix layers (QSLs, \citealt{titov2002}) where the magnetic field line linkage changes drastically in the volume. The strength of QSLs is measured by squashing factor Q. From the simulated magnetic structure, we computed Q with the procedures of \citet{LiuRui2016} and its map is displayed in Figure~\ref{fig_qf_2d}a. In this map, QSLs with large Q values ($>10^6$) are identified by intense white traces in strong field regions. Continuous trace of Q can be noticed along the main PIL. 

To compare QSL locations with the flare ribbon emission, the AIA 1600~\AA, ~304\AA, 131~\AA~observations during the impulsive phase of the flare are displayed in the panels of Figure~\ref{fig_qf_2d}(b-d) respectively. On these images, contours of Q at $10^5$, $10^6$ levels are over-plotted by removing the irrelevant QSLs with a mask on the Q-map. The ribbon emission is related to the erupting MFR in the SW region, which is co-spatial with the QSL section of the simulated magnetic configuration. Unlike the QSLs determined in the NLFFF models \citep{Vemareddy2021_MagStru}, the QSLs in this simulation better represent the twisted core flux along the PIL and the observed ribbon emission. 

\begin{figure*}[!ht]
    \centering
    \includegraphics[width=0.9\textwidth]{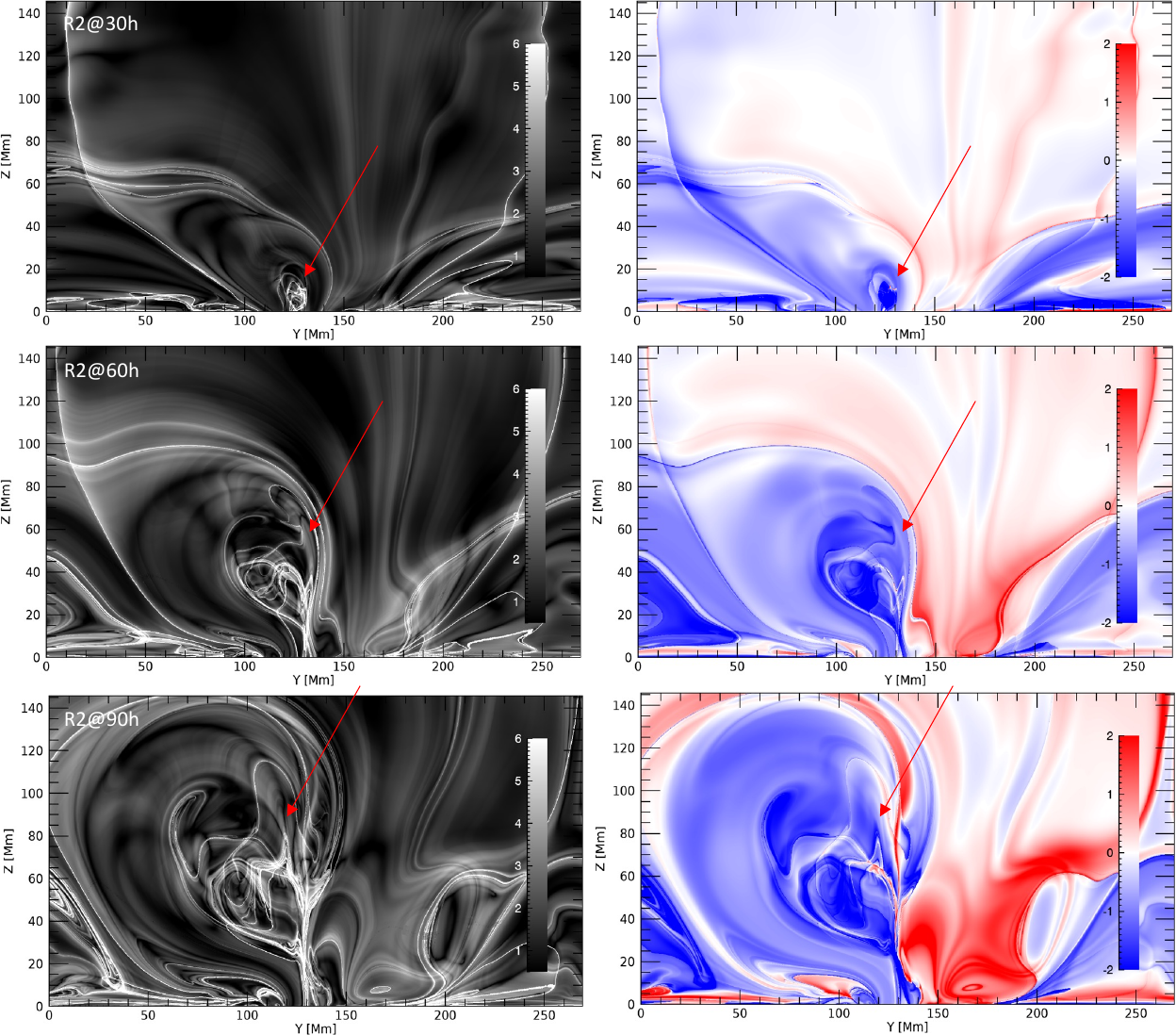}
    \caption{$ln(Q)$ and twist distribution in the vertical slice placed across the rising flux rope as shown in Figure~\ref{fig_bemiss} {\bf Left column:} Maps of $ln(Q)$ at three different times of R2. QSLs of large Q-values (white) are boundaries of magnetic domains enclosing the twisted flux (red arrows) about the PIL. {\bf Right column:} twist distribution at three different times. Twist values refer to left handed helicalness of field lines in the flux rope. Note that the upward rise motion of the flux rope in time (top to bottom). }
    \label{fig_qtw}
\end{figure*}

From the theoretical studies, it was predicted that the flare ribbons are the photospheric/chromospheric foot prints of QSLs that encloses a twisted FR \citep{demoulin1996}. The extremities of the ribbons are found to be hook shaped for weakly twisted FRs and are spiral shaped for highly twisted FRs \citep{zhaoj2016}. The observed ribbon morphology in our case delineates inverse-S shape with co-spatial hooks in the extreme ends, therefore the enclosed flux rope is indicated to be moderately twisted. This finding is consistent with previously reported pre-eruptive NLFFF configurations \citep{bobra2008,suyingna2013, LiuRui2016, Vemareddy2021_MagStru}.  

In Figure~\ref{fig_qtw}, we display Q and twist number ($tw$) maps computed from the vertical cross section of the erupting flux rope at three different times. In these maps, the QSLs of large Q values well distinguish two closed domains belonging to twisted core sections above the PIL and the surrounding less sheared arcade. With the twisted flux (flux rope) at the core, the QSLs in the cross section resembles an inverse tear-drop shape as evidenced by the theoretical/numerical models \citep{JiangChaowei2018, vemareddy2019_VeryFast}. The twist of the field lines at the core is negative; therefore, field lines are left helical and the flux rope has an inverse-S morphology. Within the flux rope cross-section, the total twist of the field lines varies in the range of 2 turns; however, average twist number is well below 1 turn, so that the twisted flux is in stable equilibrium.  After the formation in the first few hours, the twisted flux rope expands and rises up in time, as delineated in Figure~\ref{fig_qtw}. 

\section{Summary and Discussion}
\label{Summ}
In this study, the evolution of the magnetic structure of AR 11429 is simulated by a time-dependent MF model. The computational procedure is implemented in PENCIL code, which is based on a uniform grid. Invoking the ad-hoc assumptions, the non-inductive electric fields are derived from time sequence vector magnetic field observations of the AR obtained by HMI and are supplemented to inductive components in order to inject sufficient free magnetic energy from the bottom of the computational domain, as suggested by \citet{Cheung2012_MF}. Unlike staggered grid based MHD codes \citep{Hayashi2019, Pomoell2019_EleFld}, in our implementation, the electric fields are defined at the bottom boundary, where magnetic fields are defined too, and the vertical component of the photospheric magnetic field is reproduced accurately through the induction equation (see Figure~\ref{fig_bz_corr}).

The simulated magnetic structure evolves from potential to twisted fields over the course of two days, depicting the formation of flux rope and its eventual eruption as a large-scale CME. While the ARs have a non-uniform field line twist distribution in and around the PIL, the simulated magnetic structure of R2 (driven by electric field based on $U$-assumption) is better comparable with the observed coronal morphology than R3 (driven by electric fields based on $\Omega$-assumption). The magnetic configuration is a inverse S-sigmoidal structure, composed of potential field enveloping the inverse J-shaped fields that are shared past one another and a low lying twisted field along the major PIL. Proxy emission maps are generated from the simulated field and compared with the observations. These emission maps exhibit a remarkable one-to-one correspondence with the spatial characteristics in the 304 and 335~\AA~waveband coronal pictures (Ref Fig~\ref{fig_bemiss}), particularly the filament channel that is being erupted from the SW region \citep{DhakalS2020}. We further conducted topological analysis of simulated fields to understand the validity of the modelled twisted flux rope. The QSLs are distributed co-spatially with the observed flare ribbons, consistent with the standard flare models therefore, the extent of the twist and orientation of the erupting flux rope are indicated to be consistent with the real scenario in this case. 

Employing the TMF model, we simulate the long-term coronal field evolution, correctly capturing the formation of the twisted flux rope mimicking coronal sigmoid and then its eruption as seen in the AR. These simulations are very useful to shed light on the mechanisms responsible for the CME eruptions in the source ARs. One drawback of these simulations is that the erupting phase of the flux rope is not eventual such as the observed solar cases. The eruption phase is slow as the simulation proceeds in time (see the animation associated with Figure~\ref{fig_R2_persp} ) after the twisted flux rope formed,  which is an inherent property of the MF model because the velocity is controlled by the frictional parameter. Under these circumstances, one can use the twisted field generated from these TMF simulations as the initial condition for the full MHD or data-inspired simulations (e.g., \citealt{Inoue2023_EvoEru, JiangChaowei2018}) to reproduce the actual eruption scenario,  which will be focused in our future investigations. \\

%\begin{Acknowledgements}  
SDO is a mission of NASA's Living With a Star Program. P.V. acknowledges the support from DST through the Startup Research Grant. The author benefited from attending the NORDITA workshop on solar helicities held in 2019.  Field line rendering is due to VAPOR visualization software (\url{www.vapor.ucar.edu/}). I thank the referee for a careful review providing constructive comments and suggestions which improved the presentation of the results drastically.
%\end{Acknowledgements}

%\bibliographystyle{apj}
\bibliographystyle{aasjournal}
%\bibliography{ref_lib20230703}

\end{document}